\UseRawInputEncoding 
\documentclass[preprint,preprintnumbers,amsmath,amssymb]{article}

\usepackage{graphicx}
\usepackage{dcolumn}
\usepackage{bm}
\usepackage{color}
\usepackage{lscape}
\usepackage{amscd,amsmath,amssymb}
\newcommand{\be}{\begin{equation}}
\newcommand{\bel}{\begin{equation}\label}
\newcommand{\ee}{\end{equation}}

\newlength{\intwidth}
\DeclareRobustCommand{\fpint}[2]
   {\mathop{%
      \text{%
        \settowidth{\intwidth}{$\int$}%
        \makebox[0pt][l]{\makebox[\intwidth]{$-$}}%
        $\int_{#1}^{#2}$}}}

\def\XXint#1#2#3{{\setbox0=\hbox{$#1{#2#3}{\int}$ }
\vcenter{\hbox{$#2#3$ }}\kern-.5\wd0}}

\begin{document}


\title{Generalized incompressible fluid dynamical system interpolating\\
  between the Navier-Stokes and Burgers equations\\
  in two dimensions}


\author{Koji Ohkitani\\
Research Institute for Mathematical Sciences,\\
Kyoto University, Kyoto 606-8502 Japan.}    

%
\date{\today}
\maketitle

\begin{abstract}
  We propose a set of generalized incompressible fluid dynamical equations,
  which interpolates between the Burgers and Navier-Stokes equations  in two dimensions
and study their properties theoretically and numerically.  It is well-known that
 under the assumption of potential flows the multi-dimensional Burgers equations are integrable
in the sense they can be reduced to the heat equation, via the so-called Cole-Hopf linearization.
On the other hand, it is believed that the Navier-Stokes equations do not possess such a nice property.

Take, for example,  the 2D Navier-Stokes equations and rotate the velocity gradient by 90 degrees,
we then obtain a system which is equivalent to the Burgers equations.  Based on this observation,
we introduce a system of generalized incompressible fluid dynamical equations
by rotating velocity gradient  through a continuous angle parameter. That way
we can compare properties of an integrable system with those of non-integrable ones by relating them
through a continuous parameter. Using direct numerical experiments we show how the flow properties
change (actually, deteriorate)
when we increase the angle parameter $\alpha$ from 0 to $\pi/2$. It should be noted that the case associated with least
regularity, namely the Burgers equations ($\alpha=\pi/2$),  is integrable via the heat kernel.  

We also formalize a perturbative treatment of the problem, which in principle yields the solution
of the Navier-Stokes equations  on the basis of that of the Burgers equations. The principal
variations around the Burgers equations are computed numerically and are shown to agree well with
the results of direct numerical simulations for some time.

\end{abstract}


\section{Introduction}
Some of parabolic partial differential equations (PDEs, hereafter)  are better understood
than others and those that are integrable via heat flows are typical examples of 'good'
systems. We will study the  Burgers and Navier-Stokes  in two spatial dimensions in the paper.
The former is a system for a compressible fluid and the latter for an incompressible fluid,
both of which have physical significance in their own right.
The Burgers equations for potential flows can be reduced to the heat equation
via the Cole-Hopf linearisation and are hence integrable in this sense.
In this paper we attempt to connect the well-understood
PDEs with lesser-understood ones for incompressible fluid flows.

The objective of the paper is to propose a set of generalized incompressible fluid dynamical equations
and  study their properties theoretically and numerically. The rationale is to seek a method of approximating
the 2D Navier-Stokes solutions on the basis of the 2D Burgers solutions (or, eventually the heat flow).

There are papers which compare the two equations, theoretically
\cite{KL1957, PR2016} and numerically \cite{OD2012}. To our knowledge,
we are unaware of articles which directly connect the two equations continuously
with a transformation, even in two dimensions. 
As a similar work, in \cite{LR2009} the authors study blowup of solutions to
the generalized surface quasi-geostrophic equation when the spatial gradient is rotated.
Their motivation is thus different from ours in that they focus singularity formation. 

In this paper, specifically we will {\it interpolate} between the Burgers equations and Navier-Stokes equations
in two dimensions  with  a continuous parameter, thereby seeking to improve the understanding
of the latter. The latter system is known to be well-posed, but the properties of their solutions
leave much room
for understanding, particularly when the Reynolds number is large.
The rest of this paper is organized as follows. In Section II we set down mathematical formulation.
In Section III we present numerical experiments. We discuss
Poincar{\'e}'s variational equations and self-similar solutions in Section IV.
Section V is devoted to summary and discussion.

\section{Mathematical Formulation}
\subsection{Governing equations}
We consider the incompressible 2D Navier-Stokes equations under periodic boundary conditions
on $\mathbb{T}^2=[0,2\pi]^2$
  \begin{eqnarray} 
  \frac{\partial\bm{u}}{\partial t}+\mathbb{P}(\bm{u}\cdot\nabla\bm{u})&=&\nu \triangle \bm{u},\label{2DNS}\\
  \nabla \cdot \bm{u}&=&0, \nonumber 
\end{eqnarray} 
  where  $\bm{u}(\bm{x},0)=\bm{u}_0(\bm{x})$ denotes an initial velocity and $\mathbb{P},\;
  P_{ij}=\delta_{ij}-\partial_i \partial_j \triangle^{-1},\;(i,j,=1,2)$ the solenoidal projection.
Also, consider  the 2D Burgers equations 
\bel{2DBurg}
\frac{\partial\bm{w}}{\partial t}+\bm{w}\cdot\nabla\bm{w}=\nu \triangle \bm{w},
\ee
$$\bm{w}(\bm{x},0)=\bm{w}_0(\bm{x}).$$

The following is a trivial remark on the relationship between the Navier-Stokes and Burgers equations,
but is best stated here before proceeding to their comparison.
Rewriting (\ref{2DNS}) as
$$\frac{\partial\bm{u}}{\partial t}=\mathbb{P}\left(-\bm{u}\cdot\nabla\bm{u}+\nu \triangle \bm{u}\right),$$
we observe that
{\it for short time evolution} solutions to the  Navier-Stokes equations are just a solenoidal restriction
of those to the Burgers equations. Hence,  for short time the regularity of the former follows from that of the latter.
On the other hand, for long time evolution it cannot be concluded from this observation
that such a correspondence still holds or not. Particularly in three dimensions  this is the well-known open problem.

Define the velocity potential $\phi$ by $\bm{w}=\nabla \phi$ and it satisfies the Hamilton-Jacobi equation
of the form
\bel{HJ}
\frac{\partial \phi}{\partial t}+\frac{|\nabla  \phi|^2}{2}=\nu\triangle \phi.
\ee
Now consider a new velocity field  which has $\phi$ as its stream function, that is,
$\widetilde{\bm{u}}=\nabla^\perp\phi$ so that $\nabla \cdot \widetilde{\bm{u}}=0,$ where
$\nabla^\perp=(-\partial_y,\partial_x)^T$  denotes a skewed gradient  with matrix transpose $T$.
We then have
$$\frac{\partial \nabla^\perp\phi}{\partial t}+\nabla^\perp\frac{|\nabla\phi|^2}{2}=\nu\triangle \nabla^\perp\phi.$$
Noting that
$\nabla^\perp\frac{|\nabla\phi|^2}{2}=\nabla^\perp\frac{|\nabla^\perp\phi|^2}{2}
  =\nabla^\perp\frac{|\widetilde{\bm{u}}|^2}{2}=\widetilde{\bm{u}}\cdot \nabla^\perp \widetilde{\bm{u}},$ we may write 
  $$\widetilde{\bm{u}}_t+\widetilde{\bm{u}}\cdot\nabla^{\perp}\widetilde{\bm{u}}=\nu \triangle \widetilde{\bm{u}},$$
or, equivalently,
\bel{equivBurg}
\widetilde{\bm{u}}_t+\mathbb{P}(\widetilde{\bm{u}}\cdot\nabla^{\perp}\widetilde{\bm{u}})=\nu \triangle \widetilde{\bm{u}}.
\ee
In (\ref{equivBurg}) the application of the solenoidal projection $\mathbb{P}$ is actually redundant, but we retain it since
we will need it for generalizing the system below.
We call this the equivalent Burgers equations (\ref{equivBurg})
because it has the stream function that is identical to the velocity potential
for (\ref{HJ}).

\subsection{Interpolating between 2D Navier-Stokes and Burgers equations}
\subsubsection{Velocity formulation}
On the basis  of the equivalent Burgers equation (\ref{equivBurg}), we introduce a generalized system interpolating
between the Navier-Stokes and Burgers equations.
From here on, we will write $\bm{v}$ for  $\widetilde{\bm{u}}$ and $\psi$
for $\phi$. We then consider a generalized equation for $\bm{v}=\nabla^{\perp}\psi$
\bel{gen_eq}
\bm{v}_t+\mathbb{P}(\bm{v}\cdot\nabla^{\alpha}\bm{v})=\nu \triangle \bm{v},
\ee
where  $\nabla^{\alpha}=R^{\alpha}\nabla$ denotes a  gradient operator tilted by
a rotation matrix $R^{\alpha}$ through an angle of $\alpha \;(0 \leq \alpha \leq \pi/2),$
see below for its explicit formula.
It is clear that $\alpha=0$ corresponds to the Navier-Stokes equations and $\alpha=\pi/2$ to the Burgers equations.
The system (\ref{gen_eq}) is the generalized  equation that we will study in this paper.

It is in order to write down a form of the energy budget for the case of the Burgers equation ($\alpha=\pi/2$)
\bel{budget}
\frac{d}{dt}\int \frac{|\bm{v}|^2}{2}d\bm{x}
=\frac{1}{2}\int \omega |\bm{v}|^2 d\bm{x}
-\nu \sum_{i,j=1,2}\int \left(\frac{\partial^2 \psi}{\partial x_i \partial x_j}\right)^2
d\bm{x}, 
\ee
where $\omega=\triangle \psi.$ This shows that  the energy as defined by $\int \frac{|\bm{v}|^2}{2}d\bm{x}$
is {\it not} conserved even when $\nu=0$. See Appendix A for detailed discussion.

The following note is virtually trivial, but best stated here.
If it {\it were} not for the solenoidal projection $\mathbb{P}$ in (\ref{gen_eq}), that is,
$$\bm{v}_t+ \bm{v}\cdot\nabla^{\alpha}\bm{v}=\nu \triangle \bm{v},$$
then its solution would be given by a pullback simply as $\bm{v}=\bm{w}(R^\alpha\bm{x},t),$
where $\bm{w}(\bm{x},t)$ solves the standard Burgers equations (\ref{2DBurg}).

In fact, putting  $\bm{v}=\bm{w}(\bm{y},t)$ with $\bm{y}=R^\alpha (\bm{x})$,
we have $\bm{v}_t=\bm{w}_t(\bm{y},t),$ and
$$\partial^\alpha_j v_i=R^\alpha_{jk}\frac{\partial w_i(R^\alpha \bm{x})}{\partial x_k}
=R^\alpha_{jk} R^\alpha_{lk} \frac{\partial w_i}{\partial y_l}(\bm{y})
=\frac{\partial w_i}{\partial y_j}(\bm{y})$$
as $R^\alpha_{jk} R^\alpha_{lk}=R^\alpha_{jk} R^{-\alpha}_{kl}=\delta_{jl}.$
Also by $\nabla^\alpha\cdot\nabla^\alpha=\triangle$, we have
$\triangle_{\bm{x}} \bm{v}=\triangle_{\bm{y}} \bm{w}$.
Hence, in $y$, $\bm{w}$ just solves the Burgers equations.
Such a simple trick does not work for (\ref{gen_eq}), due to extra spatial gradients contained in
$\mathbb{P}$.

\subsubsection{Stream function formulation}
Before discussing numerical computations,
it is instructive to recast the same equations in terms of the stream function $\psi$
in order to shed light on how the dynamics of (\ref{gen_eq}) changes depending on $\alpha$ .
We begin by recalling that the 2D Navier-Stokes equations can be written in terms of the stream function \cite{KO2008,KO2015} 
$$
\psi_t+ T[\nabla\psi,\nabla\psi] =\nu \triangle \psi,
$$
where (in the case of whole plane $\mathbb{R}^2$)
$$T[\nabla\psi,\nabla\psi]\equiv
\frac{1}{\pi} \fpint{\mathbb{R}^2}{}
\frac{\left[ (\bm{x}-\bm{y}) \times \nabla \psi  (\bm{y}) \right]
(\bm{x}-\bm{y}) \cdot \nabla \psi  (\bm{y})}
     {|\bm{x}-\bm{y}|^4}\;{\rm d}\bm{y}.$$         
Here, $\fpint{}{}$ denotes a principal value integral.
Actually, the generalized equations in $\psi$  reads
\bel{psi_gen}
\psi_t+\sin\alpha\,\frac{|\nabla\psi|^2}{2}
+\cos\alpha\, T[\nabla\psi,\nabla\psi] =\nu \triangle \psi.
\ee
{\bf Derivation of (\ref{psi_gen}).}\\
Taking $\nabla^\perp \cdot $ of (\ref{gen_eq}),  we have
  $$\nabla^{\perp}\cdot\bm{v}_t+\nabla^{\perp}\cdot\mathbb{P}(\bm{v}\cdot\nabla^{\alpha}\bm{v})
  =\nu \triangle \nabla^{\perp}\cdot\bm{v}.$$
Remembering $\bm{v}=\nabla^{\perp}\psi,\;\omega=\nabla^{\perp}\cdot\bm{v}=\triangle \psi,$ we get
\bel{vorticity_eq}
\omega_t+\nabla^{\perp}\cdot(\bm{v}\cdot\nabla^{\alpha}\bm{v})                     
=\nu \triangle \omega.
\ee
Inverting the Laplacian $\triangle$ and dropping an unimportant constant of integration,  we obtain
\bel{psi_eq}
\psi_t+\triangle^{-1}\nabla^{\perp}\cdot(\bm{v}\cdot\nabla^{\alpha}\bm{v})                     
=\nu \triangle \psi.
\ee
The rotated gradient has the following expression
  $$\nabla^\alpha=R(\alpha)\nabla$$
  $$=\begin{pmatrix}
  \cos \alpha & -\sin \alpha\\
  \sin \alpha & \cos \alpha\\
  \end{pmatrix}
  \begin{pmatrix}
  \partial_1 \\ \partial_2
  \end{pmatrix}
=  \begin{pmatrix}
  \cos \alpha \partial_1-\sin \alpha \partial_2\\
  \sin \alpha  \partial_1+\cos \alpha \partial_2
  \end{pmatrix}.
$$
By direct computations we find
\begin{eqnarray}
\nabla^{\perp}\cdot(\bm{v}\cdot\nabla^{\alpha}\bm{v})&=&
-\partial_2\left\{ (v_1 \cos\alpha+v_2 \sin\alpha)\partial_1 v_1
+(-v_1 \sin\alpha+v_2 \cos\alpha)\partial_2 v_1 \right\}\nonumber\\
&+&\partial_1\left\{ (v_1 \cos\alpha+v_2 \sin\alpha)\partial_1 v_2
+(-v_1 \sin\alpha+v_2 \cos\alpha)\partial_2 v_2\right\} \nonumber\\
&=&
-\left\{ (\partial_2 v_1 \cos\alpha+\partial_2 v_2 \sin\alpha)\partial_1 v_1
+(-\partial_2 v_1 \sin\alpha+\partial_2 v_2 \cos\alpha)\partial_2 v_1 \right\}\nonumber\\
&+&\left\{ (\partial_1 v_1 \cos\alpha+\partial_1 v_2 \sin\alpha)\partial_1 v_2
+(-\partial_1 v_1 \sin\alpha+\partial_1 v_2 \cos\alpha)\partial_2 v_2\right\} \nonumber\\
&-&\left\{ (v_1 \cos\alpha+v_2 \sin\alpha)\partial_1 \partial_2 v_1
+(-v_1 \sin\alpha+v_2 \cos\alpha)\partial_2~2 v_1 \right\}\nonumber\\
&+&\left\{ (v_1 \cos\alpha+v_2 \sin\alpha)\partial_1^2 v_2
+(-v_1 \sin\alpha+v_2 \cos\alpha)\partial_1\partial_2 v_2\right\} \nonumber\\
&=&\cos\alpha (\partial_1 v_2-\partial_2 v_1) (\partial_1 v_1+\partial_2 v_2)\nonumber\\  
&+&\sin\alpha \left\{(\partial_1 v_1)^2+(\partial_2 v_1)^2+(\partial_1 v_2)^2+(\partial_2 v_2)^2 \right\} \nonumber\\   
&+&(v_1 \cos\alpha+v_2\sin \alpha) \psi_{122} +(-v_1 \sin\alpha+v_2\cos \alpha) \psi_{222} \nonumber\\  
&+&(v_1 \cos\alpha+v_2\sin \alpha) \psi_{111} +(-v_1 \sin\alpha+v_2\cos \alpha) \psi_{112} \nonumber\\
&=& \sin\alpha|\nabla\bm{v}|^2+(v_1 \cos\alpha+v_2\sin \alpha) (\triangle \psi)_1
+(-v_1 \sin\alpha+v_2\cos \alpha)(\triangle \psi)_2 \nonumber\\  
&=&\sin \alpha\,(|\nabla\bm{v}|^2+v_2\partial_1 \omega-v_1\partial_2 \omega)
+\cos \alpha \,(\bm{v}\cdot \nabla) \omega\nonumber\\
&=&\sin \alpha\,\triangle \frac{|\bm{v}|^2}{2}+\cos \alpha\, (\bm{v}\cdot \nabla) \omega. \nonumber
\end{eqnarray}
 Thus,  from (\ref{vorticity_eq}) we deduce
\bel{vorticity.gen}
 \omega_t+\sin \alpha\,\triangle \frac{|\bm{v}|^2}{2}+\cos \alpha\, (\bm{v}\cdot \nabla) \omega
 =\nu \triangle \omega. \;\;\; \square
\ee
From (\ref{psi_eq}) we likewise find
$$\psi_t+\sin \alpha\, \frac{|\bm{v}|^2}{2}+\cos \alpha\, T[\nabla \psi,\nabla \psi]
=\nu \triangle \psi.$$
In particular, in the case of the Burgers equation ($\alpha=\pi/2$) we have
$$\omega_t+\triangle \frac{|\bm{v}|^2}{2}=\nu \triangle \omega,$$
which shows that $\omega$ is not a Lagrangian invariant even if $\nu=0$.
On the other hand, putting $$\omega=-2\nu \triangle \log \theta,$$ 
we  can readily deduce
$$\theta_t=\nu \triangle \theta,$$
which may be regarded as an analogue of the Cole-Hopf transform in disguise.

It can be derived as follows. Putting
$$\omega=k \triangle \log \theta$$
for a constant $k$,
we have
$$\omega_t=k \triangle \frac{\theta_t}{\theta}$$
and
$$\triangle\omega=k \triangle \frac{\theta \triangle \theta-|\nabla \theta|^2}{\theta^2}.$$
Hence
$$\omega_t-\nu \triangle\omega
=k \triangle\left( \frac{\theta_t}{\theta} -\nu \frac{\theta \triangle \theta-|\nabla \theta|^2}{\theta^2} \right)$$
$$=k \triangle\frac{\theta_t-\nu\triangle \theta}{\theta} +k\nu\triangle \frac{|\nabla \theta|^2}{\theta^2}$$
By $\omega=k\triangle \log \theta=\triangle \psi$ and $\psi=k \log\theta,$ discarding an insignificant constant of
integration, we have $\bm{v}=\nabla^\perp\psi=k\frac{\nabla^\perp \theta}{\theta}.$
Hence we find
$$\omega_t-\nu \triangle\omega - \frac{\nu}{k} \triangle |\bm{v}|^2
=k\triangle \frac{\theta_t -\nu\triangle \theta}{\theta}.$$
Taking $k=-2\nu,$ we complete the derivation. $\square$

In passing we note that in terms of the velocity  $\bm{v}=-2\nu \nabla^{\perp} \log \theta,$
the governing equations for $\alpha=\pi/2$ can also be written
$$\bm{v}_t+\bm{v}\cdot\nabla^\perp \bm{v}=\nu \triangle \bm{v}.$$
We recall the explicit solution of $\theta$ in $\mathbb{R}^2$
$$
\theta(\bm{x},t)=\frac{1}{4\pi \nu t}\int_{\mathbb{R}^2}\theta_0(\bm{x}')
\exp\left(-\frac{|\bm{x}-\bm{x}'|^2}{4\nu t}\right)  d\bm{x}',
$$
where $\bm{v}_0=-2\nu\nabla^\perp\log\theta_0.$

Some remarks for the general cases of $\alpha$ are in order.
When $\nu >0$, the system is expected to be well-posed for $0 < \alpha < \frac{\pi}{2}$,
but its proof is not known at the moment and seems non-trivial. One reason for this is that
the maximum principle is not available unlike the 2D Burgers equation due to nonlocal interaction.
Also, enstrophy is not well-controlled unlike the 2D Navier-Stokes equations.
Therefore we present in Section IV a formal perturbative analysis, which indicates well-posedness of
the system, at least for $\alpha \lesssim \pi/2$.

\begin{table}[h]
\begin{center} 
\begin{tabular}{|c|c|c|c|}
\hline
$\alpha$   & 0 & $\ldots\ldots\ldots\ldots\ldots$ & $\pi/2$ \\ \hline
viscous flows &  $\stackrel{\mbox{regular}}{\mbox{(Navier-Stokes equations)}}$  & expected to be regular & $\stackrel{\mbox{regular}}{\mbox{(Burgers equations)}}$  \\ \hline
inviscid flows & $\stackrel{\mbox{regular}}{\mbox{(Euler equations)}}$ & expected to be singular & singular(blowup)\\
\hline
\end{tabular}
\end{center} 
\caption{Generalized incompressible equations in two dimensions}
\label{Table.1}
\end{table}

\section{Numerical experiments}
\subsection{Numerical methods}
We solve (\ref{gen_eq}) by a standard pseudo-spectral method under periodic boundary conditions
on $\mathbb{T}^2=[0,2\pi]^2$. The number of grid points used are $N^2=512^2, 1024^2$ and $2048^2.$
Time marching was performed  by the fourth-order　Runge-Kutta method, with a typical time step
$\Delta t=1\times 10^{-3}$.
As an initial condition we take
\bel{IC}
\omega_0(\bm{x})=\sin x \sin y+\cos y,
\ee
which was used in \cite{CMT1994}.  
\subsection{Numerical results}
\begin{figure}[htbp] 
\begin{minipage}{.49\linewidth}
  \includegraphics[scale=0.5,angle=0]{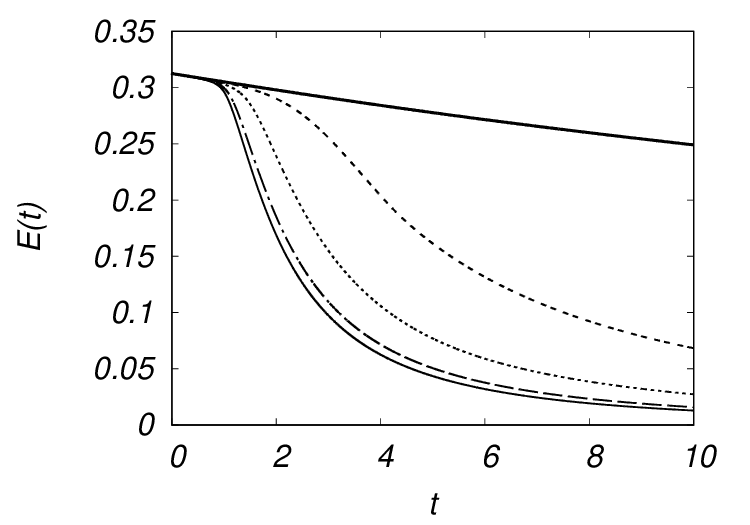}
  \caption{Time evolution of energy $E(t)$ for $\alpha=0$ (thick solid), $\pi/8$ (dashed), $\pi/4$ (dotted),
    $3\pi/8$ (dash-dotted) and $\pi/2$ (solid).}
  \label{ene}
\end{minipage}
\begin{minipage}{.49\linewidth}
  \includegraphics[scale=0.5,angle=0]{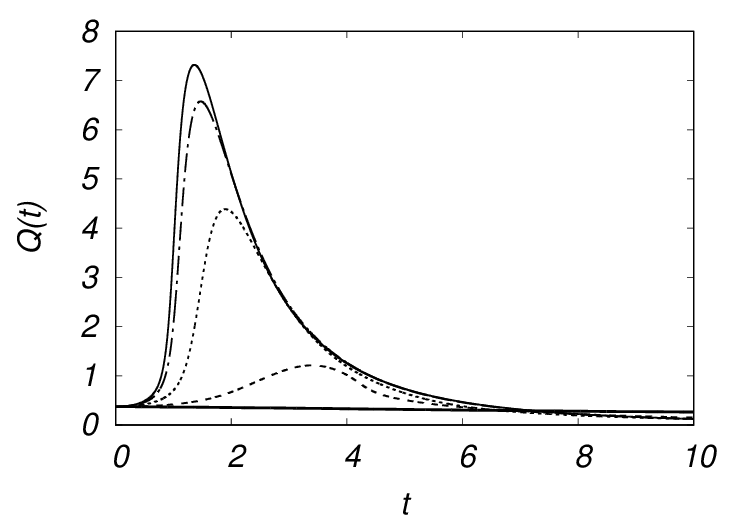}
  \caption{Time evolution of enstrophy $Q(t)$ for  $\alpha=0$ (thick solid), $\pi/8$ (dashed), $\pi/4$ (dotted),
     $3\pi/8$ (dash-dotted) and $\pi/2$ (solid).} 
  \label{ens}
\end{minipage}\end{figure}

  We begin studying the time evolution of (squared) norms, which are spatial averages of energy
  $$E(t)=\frac{1}{(2\pi)^2}\int_{\mathbb{T}^2} \frac{|\bm{v}|^2}{2} d\bm{x},$$
  enstrophy
  $$Q(t)=\frac{1}{(2\pi)^2}\int_{\mathbb{T}^2} \frac{\omega^2}{2} d\bm{x},$$
  and palinstrophy
  $$P(t)=\frac{1}{(2\pi)^2}\int_{\mathbb{T}^2} \frac{|\nabla\omega|^2}{2} d\bm{x}.$$
  In Fig.\ref{ene} we show time evolution of energy for five different values of $\alpha$.
  At $\alpha=0$ (the Navier-Stokes case)  the energy decays most slowly, virtually linear in time,
  reflecting the conservation in the limit of varnishing viscosity.
  As $\alpha$ increases, energy decay takes place  more rapidly. The case of $\alpha=3\pi/8$ is already very
  close to that of $\alpha=\pi/2$ (the Burgers case).

\begin{figure}[htbp] 
\begin{minipage}{.49\linewidth}
  \includegraphics[scale=0.6,angle=0]{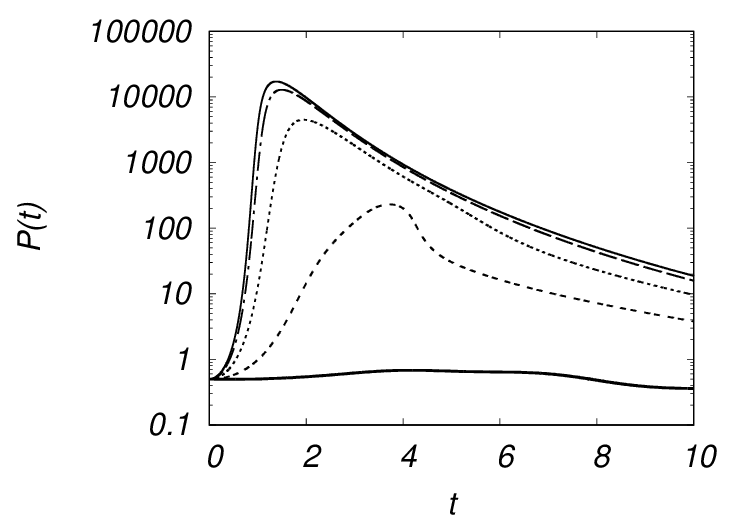}
  \caption{Time evolution of energy $P(t)$ for $\alpha=0$ (thick solid), $\pi/8$ (dashed), $\pi/4$ (dotted),
    $3\pi/8$ (dash-dotted) and $\pi/2$ (solid).}
  \label{pal}
\end{minipage}
\end{figure}

  In Fig.\ref{ens} we show time evolution of enstrophy in a similar manner.  At $\alpha=0$
  (the Navier-Stokes case)
  enstrophy is virtually constant. At $\alpha=\pi/8$ its increase is mild, but already noticeable
  and the enstrophy grows with the increase  of $\alpha,$ particularly $\alpha \geq 3\pi/8$, so does
  its peak  value.   In Fig.\ref{pal} we show time evolution of palinstrophy in a semi-logarithmic
  plot. It is observed that
  for $\alpha \geq  3\pi/8,$ $P(t)$ grows rapidly  in time, consistent with  the sustained
  nontrivial dissipation   of enstrophy.
  
\begin{figure}[htbp]
\begin{minipage}{1.\linewidth}
  \includegraphics[scale=1.,angle=0]{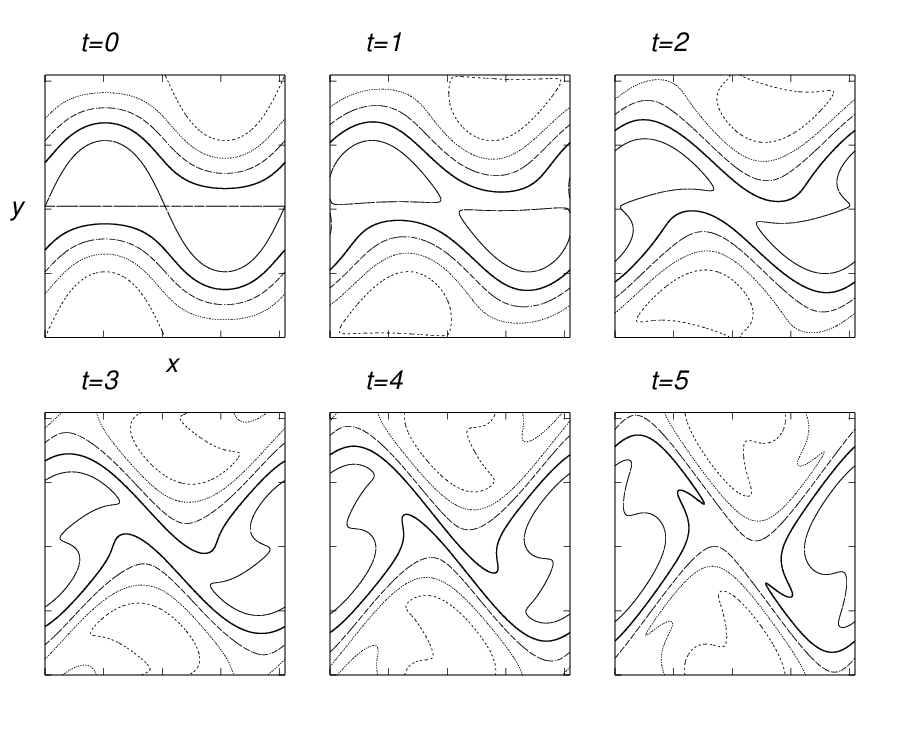}
  \caption{Contours of vorticity for $\alpha=0$ (the Navier-Stokes case), with thresholds $\omega=0, \pm 0.5, \pm 1.0, \pm 1.5,
  \pm 2.0.$}
  \label{cont_0}
\end{minipage}
\end{figure}

\begin{figure}[htbp]
\begin{minipage}{1.\linewidth}
  \includegraphics[scale=1.,angle=0]{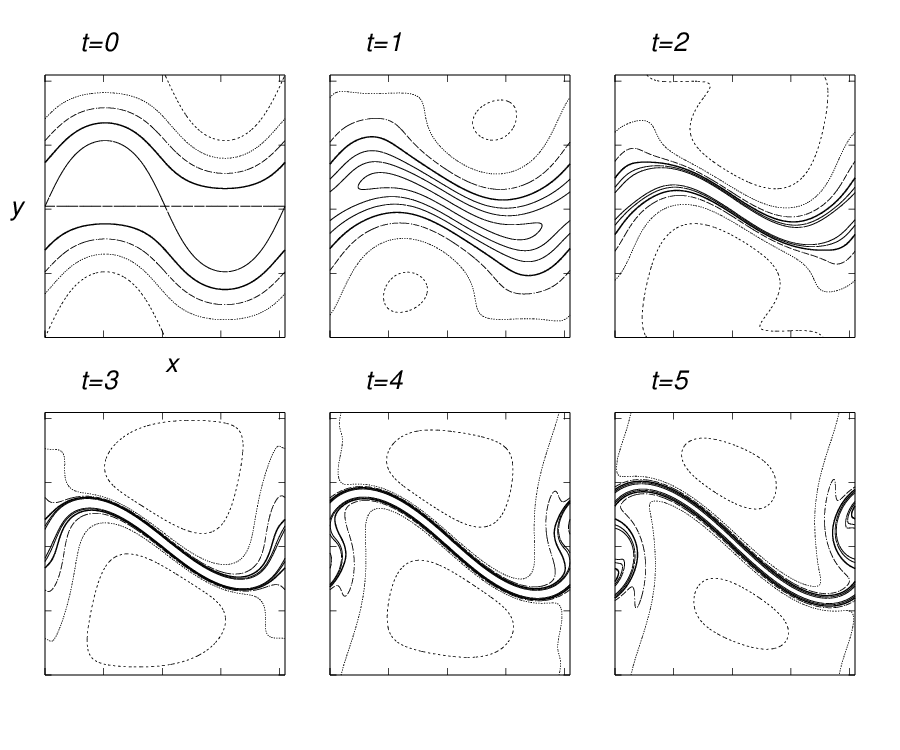}
  \caption{Contours of vorticity for $\alpha=\pi/8$ plotted as in Fig.\ref{cont_0}.}
\label{cont_pi8}
\end{minipage}
\end{figure}

We now turn to study spatial structure of the flow fields using vorticity as defined by
$\omega=\triangle \psi$ for all the values of $\alpha$.
In Fig.\ref{cont_0} we show the time evolution of contour plots of vorticity
for $\alpha=0$ (the Navier-Stokes case).
Around $t=4$, layers of sharp vorticity gradients  are formed near the center at an angle
of $\approx \pi/4$ with the $x$-axis. That is about the time  when the palinstrophy takes a mild maximum.
In Fig.\ref{cont_pi8} we show the time evolution of contour plots of vorticity for $\alpha=\pi/8$.
Formation of sharp vorticity gradient is observed more strikingly  than the case $\alpha=0$,
and the angle they make with the $x$-axis
is shallower than it is for $\alpha=0$.
At $\alpha=\pi/4$, the angle that the layers make with the $x$-axis becomes even shallower.
At $\alpha=3\pi/8$, we observe  wavy layers being formed along the $x$-axis. Actually they are very much similar to the shock formation for the Burgers equations ($\alpha=\pi/2$).

\begin{figure}[htbp]
\begin{minipage}{1.\linewidth}
  \includegraphics[scale=1.,angle=0]{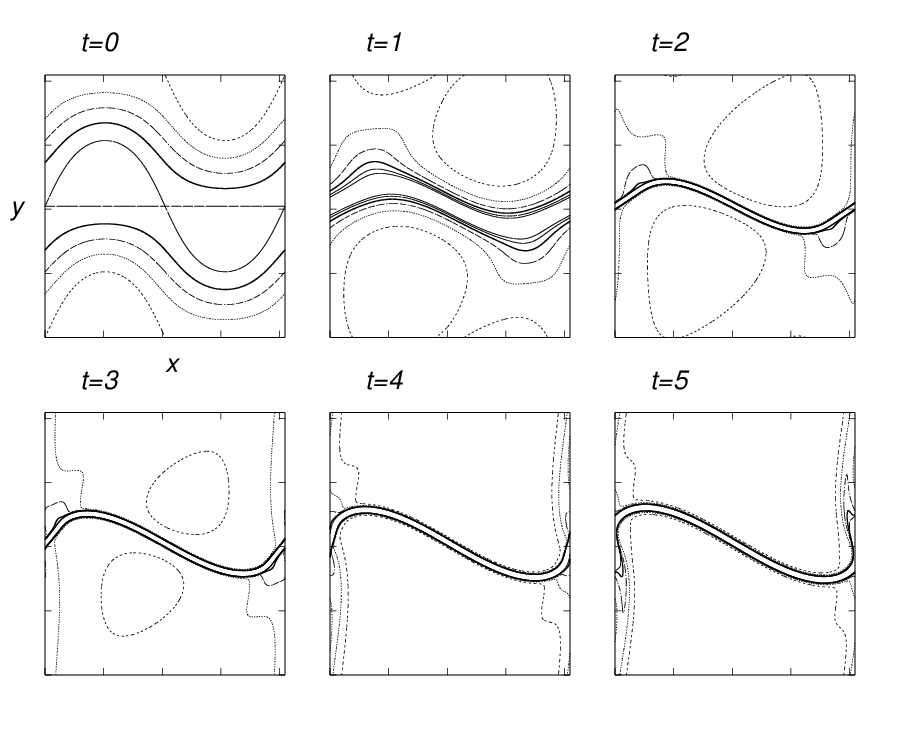}
  \caption{Contours of vorticity for $\alpha=\pi/4$ plotted as in Fig.\ref{cont_0}.}
\label{cont_pi4}
\end{minipage}\end{figure}

\begin{figure}[htbp]   
\begin{minipage}{1.\linewidth}
  \includegraphics[scale=1.,angle=0]{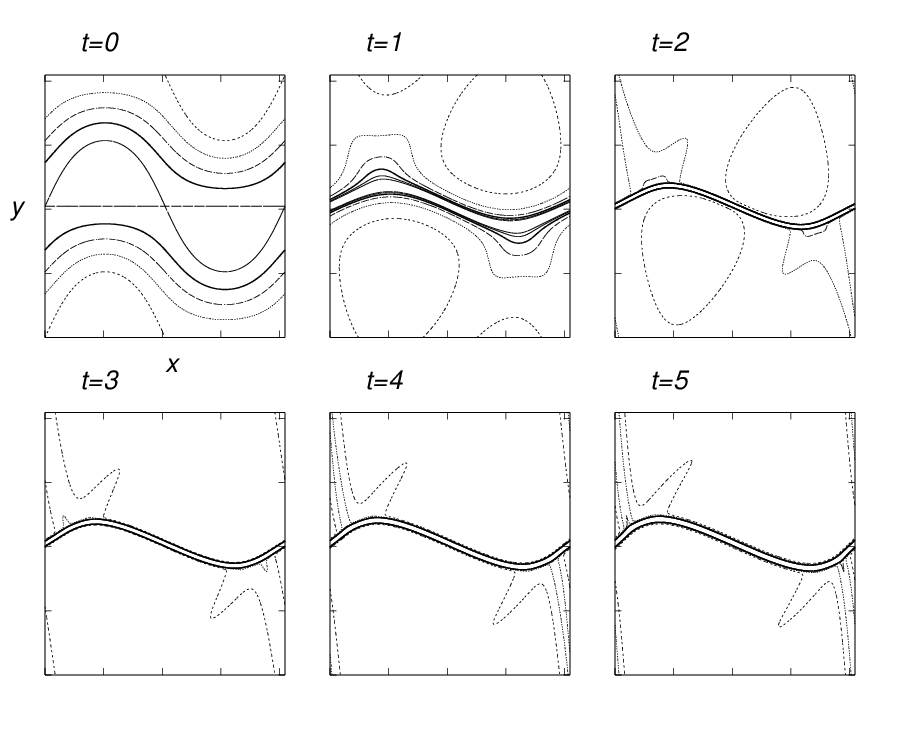}
  \caption{Contours of vorticity for $\alpha=3\pi/8$ plotted as in Fig.\ref{cont_0}.}
\label{cont_3pi8}
\end{minipage}
  \end{figure}

\begin{figure}[htbp]
  \begin{minipage}{1.\linewidth}
  \includegraphics[scale=1.,angle=0]{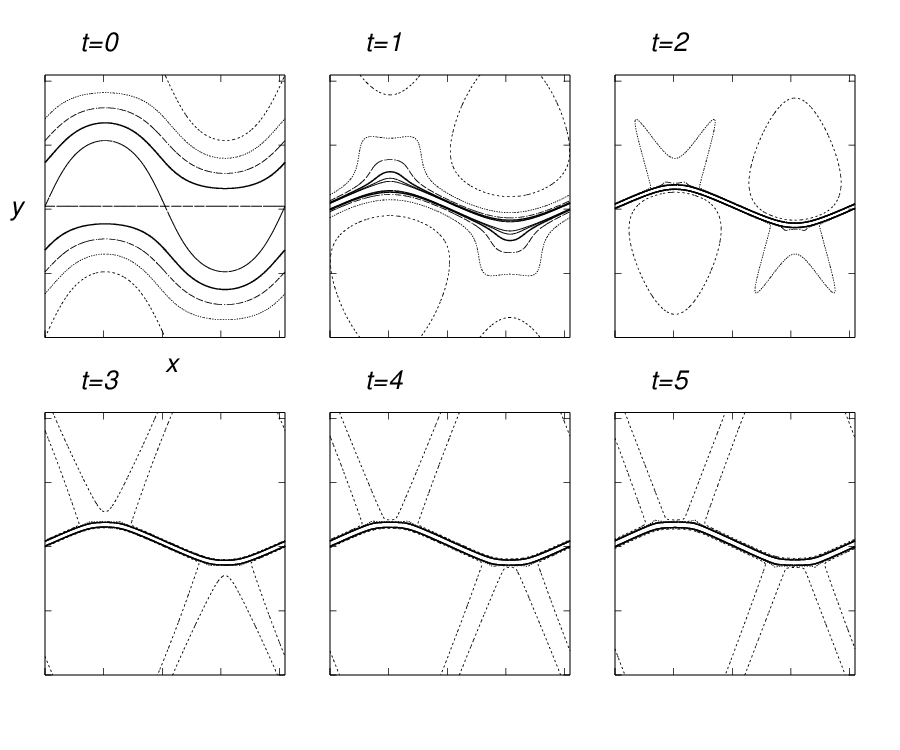}
  \caption{Contours of vorticity for $\alpha=\pi/2$ (the Burgers case) plotted as in Fig.\ref{cont_0}.}
  \label{cont_pi2}
  \end{minipage}  
\end{figure} 

\begin{figure}[htbp]
  \begin{minipage}{1.\linewidth}
  \includegraphics[scale=1.,angle=0]{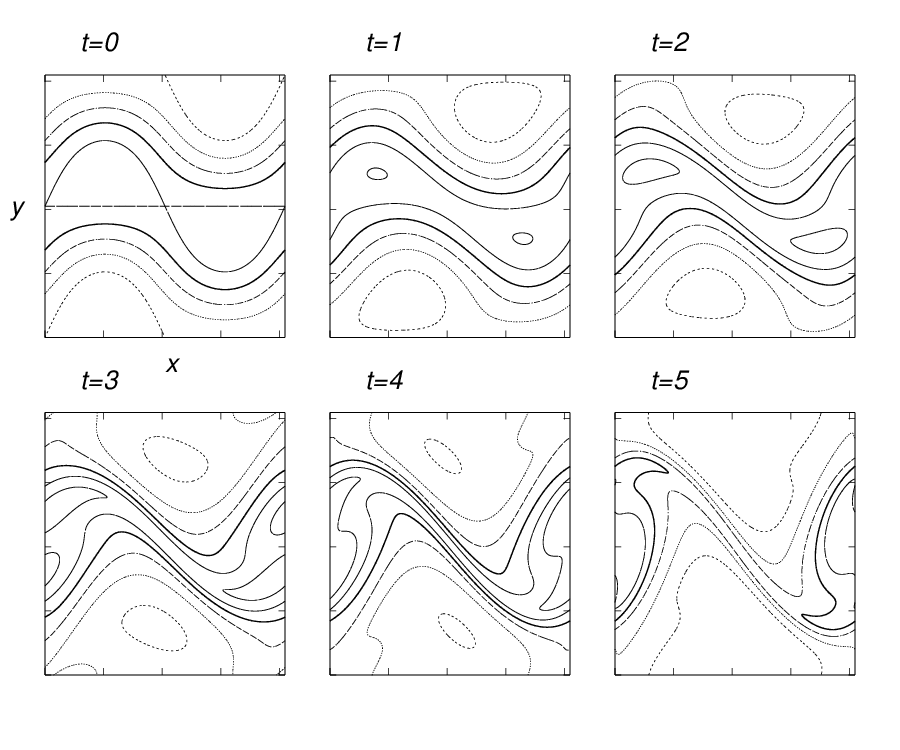}
  \caption{Contours of vorticity for $\alpha=\pi/32$ ($\gtrsim$ the Navier-Stokes case) plotted
    as in Fig.\ref{cont_0}.}
  \label{cont_pi32}
  \end{minipage}  
\end{figure} 

\begin{figure}[htbp]
  \begin{minipage}{1.\linewidth}
\includegraphics[scale=1.,angle=0]{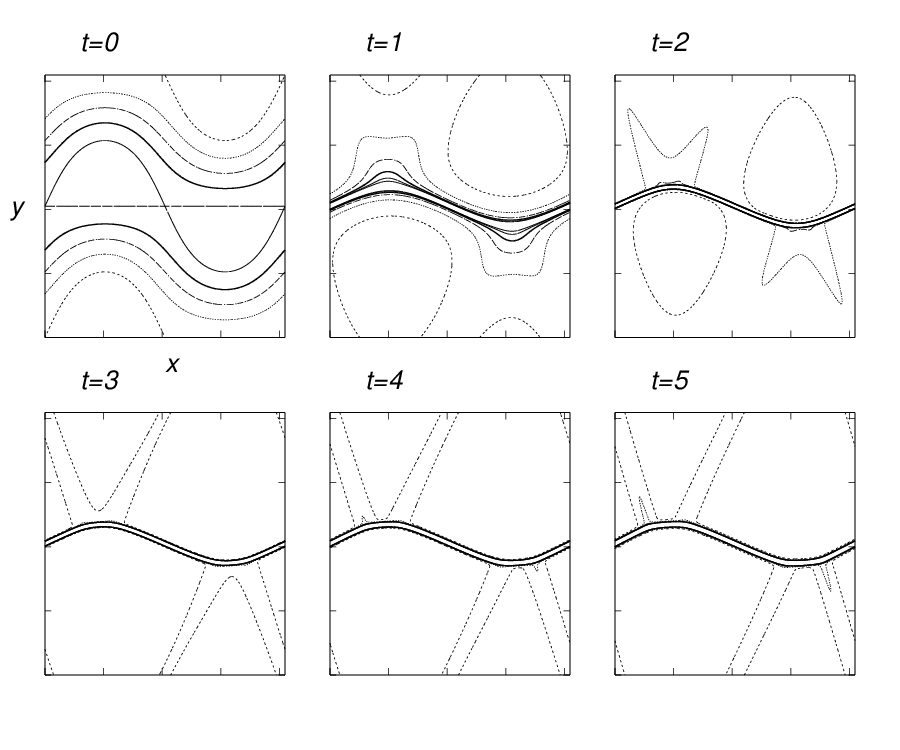}
  \caption{Contours of vorticity for $\alpha=15\pi/32$ ($\lesssim$ the Burgers case)
    plotted as in Fig.\ref{cont_0}.}
  \label{cont_15pi32}
  \end{minipage}  
\end{figure}
\begin{figure}[htbp]
  \begin{minipage}{.49\linewidth}
  \includegraphics[scale=.6,angle=0]{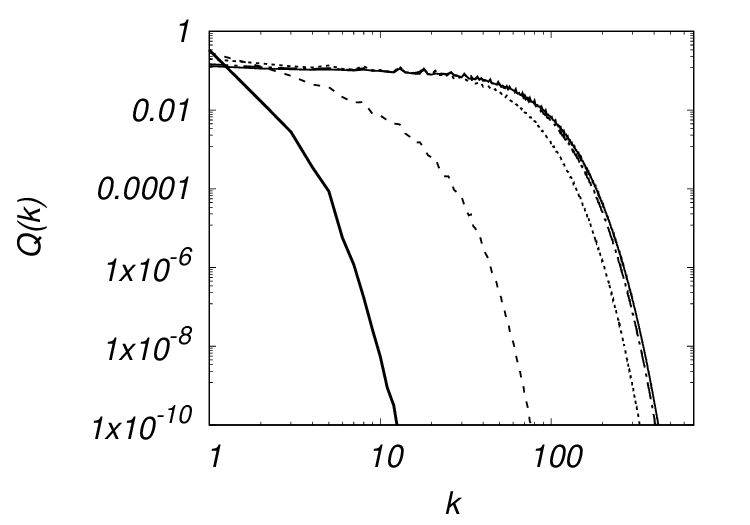}
  \caption{Enstrophy spectra at $t=2$, for $\alpha=0$ (thick solid), $\pi/8$ (dashed), $\pi/4$ (dotted),
  $3\pi/8$ (dash-dotted) and $\pi/2$ (solid).}
  \label{sp_t2}
\end{minipage}
\begin{minipage}{.49\linewidth}
  \includegraphics[scale=.6,angle=0]{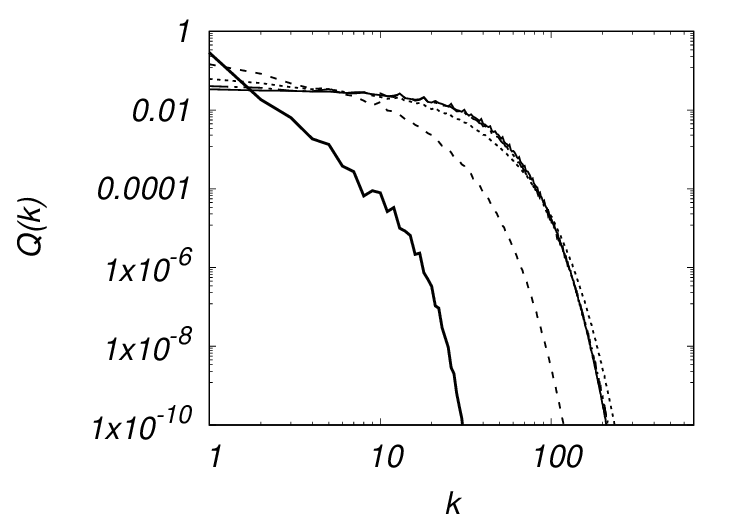}
  \caption{Enstrophy spectra at $t=5$ plotted for several values of $\alpha$,
    as in Fig.\ref{sp_t2}.}
  \label{sp_t5}
\end{minipage}
\end{figure}

\begin{figure}[htbp]
 \begin{minipage}{.49\linewidth}
  \includegraphics[scale=.6,angle=0]{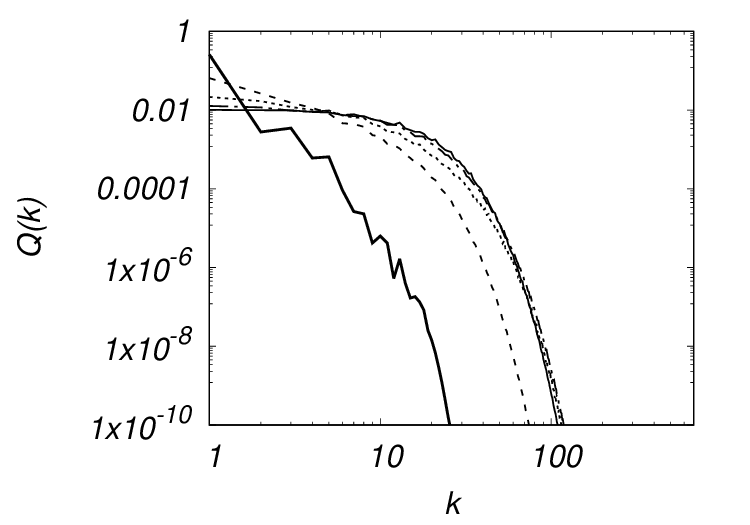}
  \caption{Enstrophy spectra at $t=10$ plotted for several values of $\alpha$,
    as in Fig.\ref{sp_t2}.}
  \label{sp_t10}
\end{minipage}
\end{figure}

 
To study the spectral properties we introduce the Fourier spectrum of enstrophy.
Defining a Fourier series expansion of vorticity by
$$\omega(\bm{x},t)=\sum_{\bm{k}} \hat{\omega}(\bm{k},t)e^{i \bm{k}\cdot\bm{x}},$$
we introduce the enstrophy spectrum $Q(k)$ as
  $$Q(k)=\frac{1}{2}\sum_{k\leq |\bm{k}| < k+1} |\hat{\omega}(\bm{k},t)|^2.$$ 
Recall that it is related to the energy spectrum $E(k)$ as $Q(k)=k^2 E(k)$.
The spectrum $Q(k)$ is composed of a few line spectra, for the initial condition (\ref{IC}).
In Fig.\ref{sp_t2} we show the enstrophy spectra at $t=2$. We observe  plateaus
in the enstrophy spectra for $\alpha=\pi/4,3\pi/8,$ and $\pi/2$,
which are consistent with the scaling $E(k) \propto k^{-2},$ associated with the shock formation for those values of $\alpha$.
At $t=5$ in Fig.\ref{sp_t5} the spectrum for $\alpha=\pi/8$  extends its excitation a bit wider, but no flat part is observed.
All the spectra have lowered their amplitudes as a result of dissipation, in comparison to Fig.\ref{sp_t5}.
In Fig.\ref{sp_t10} the flat parts of the spectra have shrunk for all the values of $\alpha$.
The case with most substantial decay is observed for $\alpha=0$ (the Navier-Stokes case),
which shows a rapid fall-off of the spectrum.

\section{Theoretical analyses}
\subsection{Variational equations}
As noted above,  the 2D Navier-Stokes  equations are not integrable
as opposed to the Burgers equations. In this section we attempt to approximate
solutions to the Navier-Stokes  equations on the basis of solutions of the Burgers equations
and {\it vice versa}.
As a first step we will consider flows near the Burgers limit ($\alpha \lesssim \pi/2$)
and also those near the Navier-Stokes limit ($\alpha \gtrsim 0$).

\subsubsection{From the Burgers limit towards the Navier-Stokes case}
We present a perturbative treatment by varying the angle parameter, on the basis of
Poincar{\'e}'s variational equations \cite{Yosida1960}, adapted here for PDEs.
It is convenient to introduce a complementary angle $\beta=\frac{\pi}{2}-\alpha$
and  consider the generalized vorticity equation
\bel{vort_beta}
\frac{\partial \omega}{\partial t} +\cos\beta\,\triangle \frac{|\bm{v}|^2}{2}+\sin \beta\,
(\bm{v}\cdot\nabla) \omega=\nu \triangle \omega
\ee
on that basis. We assume that for the solution $\left.\omega(\bm{x},t)\right|_{\beta=\beta_0}$
is known for $\beta=\beta_0$.
Using the multinomial expansion  formula for any functions $f,g,h$
$$
\partial_\beta^n (fgh)=\sum_{p+q+r=n \atop p,q,r\geq 0}\frac{n!}{p!q!r!}f^{(p)} g^{(q)} h^{(r)},
$$
where e.g. $f^{(p)}\equiv \partial_\beta^p f,$
we differentiate (\ref{vort_beta}) $n$-times  with respect to $\beta$ and put $\beta=\beta_0$.
We find
\bel{variation_B}
\frac{\partial \omega^{(n)}}{\partial t}+\sum_{p+q+r=n \atop p,q,r\geq 0}\left\{
\cos(\beta_0+p\pi/2)\triangle\frac{\bm{v}^{(q)}\cdot\bm{v}^{(r)}}{2}
+\sin(\beta_0+p\pi/2) \nabla\cdot(\bm{v}^{(q)}\omega^{(r)})\right\}=\nu \triangle \omega^{(n)},
\ee
where $\omega^{(k)}=\nabla^{\perp} \cdot \bm{v}^{(k)},\;(k=0,1,2,\ldots,n)$.
Note that $\omega^{(0)}=\left.\omega\right|_{\beta=\beta_0}, \bm{v}^{(0)}=\left.\bm{v}\right|_{\beta=\beta_0}$
denote respectively vorticity and velocity of the Burgers  equations and 
$\omega^{(1)}=\Omega, \bm{v}^{(1)}=\bm{V}$ the principal variations thereof.
Since the equations (\ref{variation_B}) are {\it linear} in $\omega^{(n)}$ or $\bm{v}^{(n)}$,
in principle we can solve them on the basis of solutions to the Burgers equations ($n=0$).
Hence we can write, at least formally,
\bel{series}
\omega(\bm{x},t)=\left.\omega(\bm{x},t)\right|_{\beta=\beta_0}
+\sum_{n=1}^{\infty}(\beta-\beta_0)^n \omega^{(n)}(\bm{x},t),  
  \ee
  for small $|\beta-\beta_0|$.
Recall that
$$\omega^{(1)} (\bm{x},t)=\lim_{\beta \to \beta_0}
\frac{\omega(\bm{x},t)-\left.\omega(\bm{x},t)\right|_{\beta=\beta_0}}{ \beta-\beta_0}$$
represents the principal variation arising from the small change of the parameter.
 As it stands, the convergence property of the series (\ref{series}) is   not known.
  
In particular, for the principal variation,
$\Omega=\dfrac{\partial \omega}{\partial \beta}$  around the Burgers equations,
differentiating (\ref{vort_beta}) we have 
\bel{principal_B}
\Omega_t-\sin \beta \,\triangle \frac{|\bm{v}|^2}{2}+\cos \beta\, (\bm{v}\cdot\nabla)\omega
+\cos \beta\, \triangle(\bm{v}\cdot\bm{V})+\sin \beta\,\nabla\cdot(\omega \bm{V}+\Omega \bm{v})
=\nu\triangle \Omega.
\ee
Setting $\beta=0,$ we find
$$\Omega_t+  (\bm{v}_{\rm B}\cdot\nabla) \omega_{\rm B}
+\triangle( \bm{v}_{\rm B}\cdot\bm{V})
=\nu\triangle \Omega,$$
where $\bm{v}_{\rm B}\equiv\left.\bm{v}\right|_{\beta=0}$ and $\omega_{\rm B}\equiv\left.\omega\right|_{\beta=0}$
respectively denote the velocity and vorticity of the Burgers equations
and $\Omega=\nabla^{\perp}\cdot\bm{V}$ the principal variation thereof
\footnote{Equivalently, one can simply put $n=1,\,\beta_0=0$ in (\ref{principal_B}).}.
Once $\Omega$ is obtained, the leading-order approximation can be constructed as
$$\omega=\omega_{\rm B}+\beta\left.\Omega\right|_{\beta=0}$$
for small $\beta\;(>0).$

Given a solution to the Burgers equations, (\ref{principal_B})  is  a system linear in the variations
$\bm{\Omega}$ and $\bm{V}$ to be solved with the initial condition $\Omega(\bm{x},0)=0.$
Since it is difficult to  obtain an analytical solution to it,
we numerically solve it as a simultaneous system with the Burgers equations.
For the Burgers equations, the same initial condition as (\ref{IC}) was used.

\begin{figure}[htbp]
\begin{minipage}{1.\linewidth}
  \includegraphics[scale=1.,angle=0]{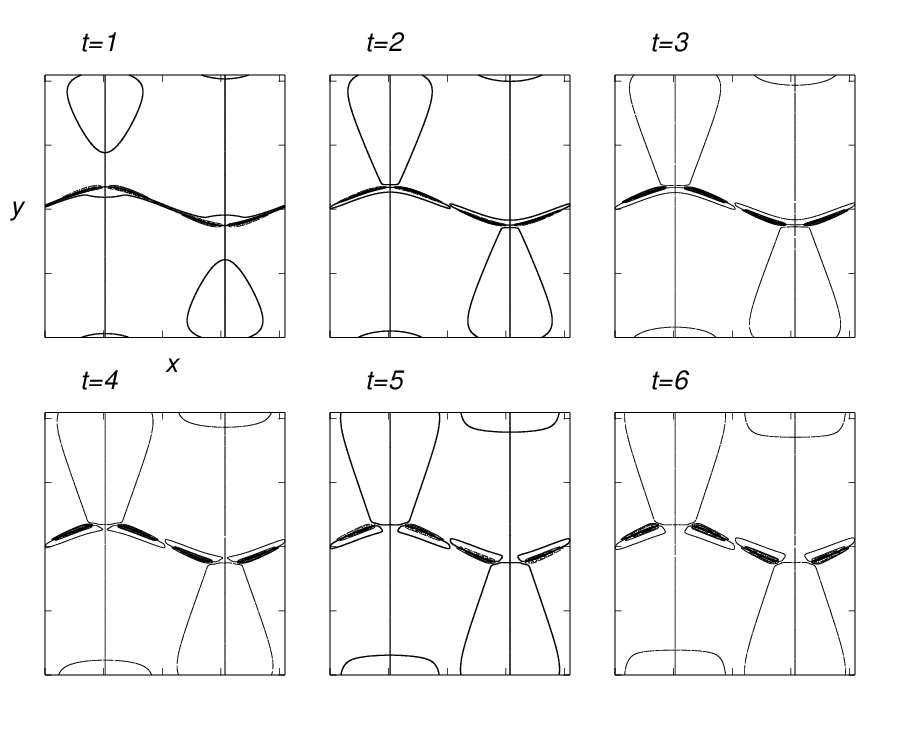}
  \caption{Vorticity contours of principal variation $\left.\Omega\right|_{\beta=0},$
    estimated at the Burgers limit. The thresholds are
    $\pm n \max_{\bm{x}}\left. \Omega(\bm{x},t)\right|_{\beta=0}/4, n=0,1,2,3,4.$}
  \label{B2N}
\end{minipage}
\end{figure}

\begin{figure}[htbp]
\begin{minipage}{1.\linewidth}
  \includegraphics[scale=1.,angle=0]{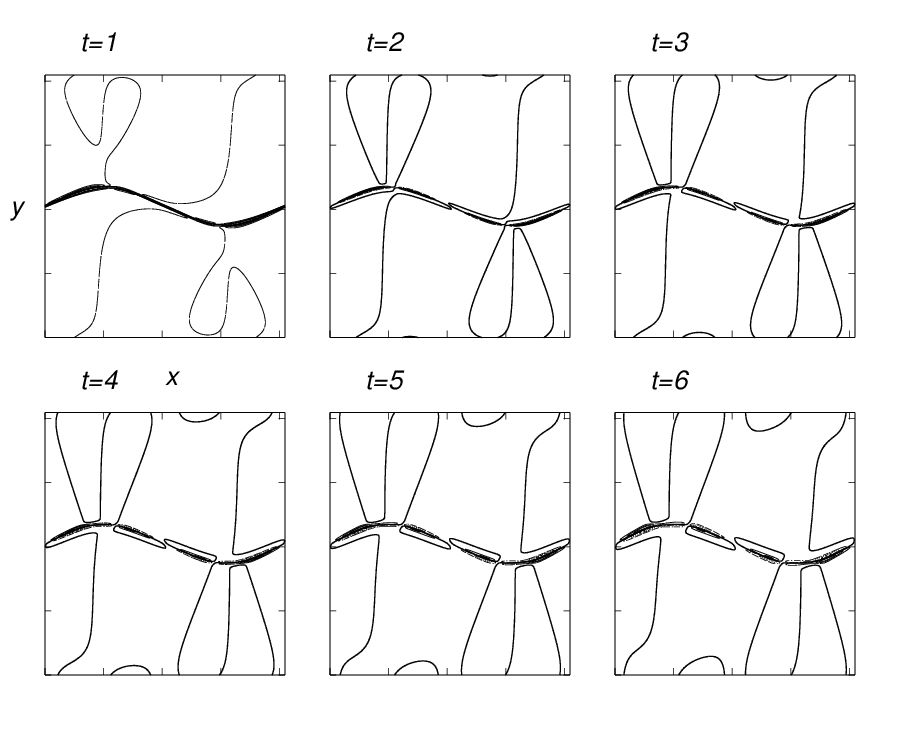}
  \caption{Contours of vorticity difference
    $\delta\omega=\left.\omega\right|_{\beta=\frac{\pi}{32}} - \omega_{\rm B}$
    near the Burgers limit, plotted as in Fig.\ref{B2N}.}
  \label{omega_diff_Bg}
\end{minipage}
\end{figure}

In Fig.\ref{B2N} we show evolution of vorticity contours of  principal variation $\left.\Omega\right|_{\beta=0}.$ 
It is clear that formation of shock layers is the dominant features throughout the computation.
To assess the performance of the variational equation, in Fig.\ref{omega_diff_Bg} we show contours of
the vorticity difference  $\delta\omega=\left.\omega\right|_{\beta=\frac{\pi}{32}} - \omega_{\rm B}$,
which look similar to those in Fig.\ref{B2N}.
In view of the relation $\delta\omega=\beta\Omega$ with $\beta=\frac{\pi}{32},$
this shows that the principal variation captures the deviation of vorticity due to the small
changes in $\beta$.

\subsubsection{From the Navier-Stokes limit  towards Burgers case}
As for the reverse direction, we restrict ourselves to the description of the leading-order
approximations, since the general formulation remains the same as before.
Differentiating
$$\omega_t +\sin\alpha\, \triangle \frac{|\bm{v}|^2}{2}+\cos \alpha\,
(\bm{v}\cdot\nabla) \omega=\nu \triangle \omega$$
with respect to $\alpha,$ we find
$$\Omega_t+\cos \alpha\, \triangle \frac{|\bm{v}|^2}{2}
-\sin \alpha\, (\bm{v}\cdot\nabla)\omega+\sin \alpha \,\triangle(\bm{v}\cdot\bm{V})
+\cos\alpha\, \nabla\cdot(\bm{V}\omega+\bm{v}\Omega) =\nu\triangle\Omega,$$
where $\Omega=\dfrac{\partial \omega}{\partial\alpha}.$
In particular, at $\alpha=0$ we get
\bel{principal_NS}
\Omega_t+ \triangle \frac{|\bm{v}_{\rm NS}|^2}{2}
+\nabla\cdot(\bm{V}\omega_{\rm NS}+\bm{v}_{\rm NS}\Omega) =\nu\triangle\Omega,
\ee
where $\bm{v}_{\rm NS} \equiv \left.\bm{v}\right|_{\alpha=0}$ and  $\omega_{\rm NS}\equiv\left.\omega\right|_{\alpha=0}$
respectively denote the velocity and vorticity of the Navier-Stokes equations
and $\Omega$ the principal variation thereof. 
The corresponding approximation can be written
$$\omega=\omega_{\rm NS}+\alpha\left.\Omega\right|_{\alpha=0}$$
for small $\alpha\;(>0).$

We  solve (\ref{principal_NS}) numerically as a simultaneous system with the Navier-Stokes equations,
with the initial condition $\Omega(\bm{x},0)=0.$
For the Navier-Stokes equations the same initial condition as (\ref{IC}) was used.

\begin{figure}[htbp]
\begin{minipage}{1.\linewidth}
  \includegraphics[scale=1.,angle=0]{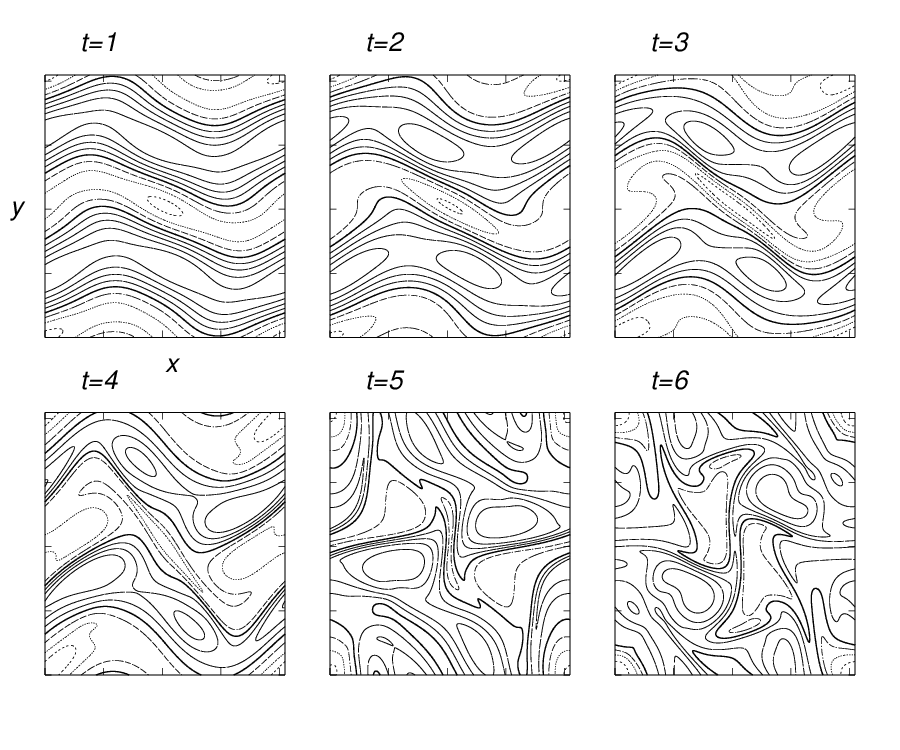}
  \caption{Vorticity contours  of principal variation $\left.\Omega\right|_{\alpha=0},$
    estimated at the Navier-Stokes limit, plotted as in Fig.\ref{B2N}.}
  \label{N2B}
\end{minipage}
\end{figure}

\begin{figure}[htbp]
\begin{minipage}{1.\linewidth}
  \includegraphics[scale=1.,angle=0]{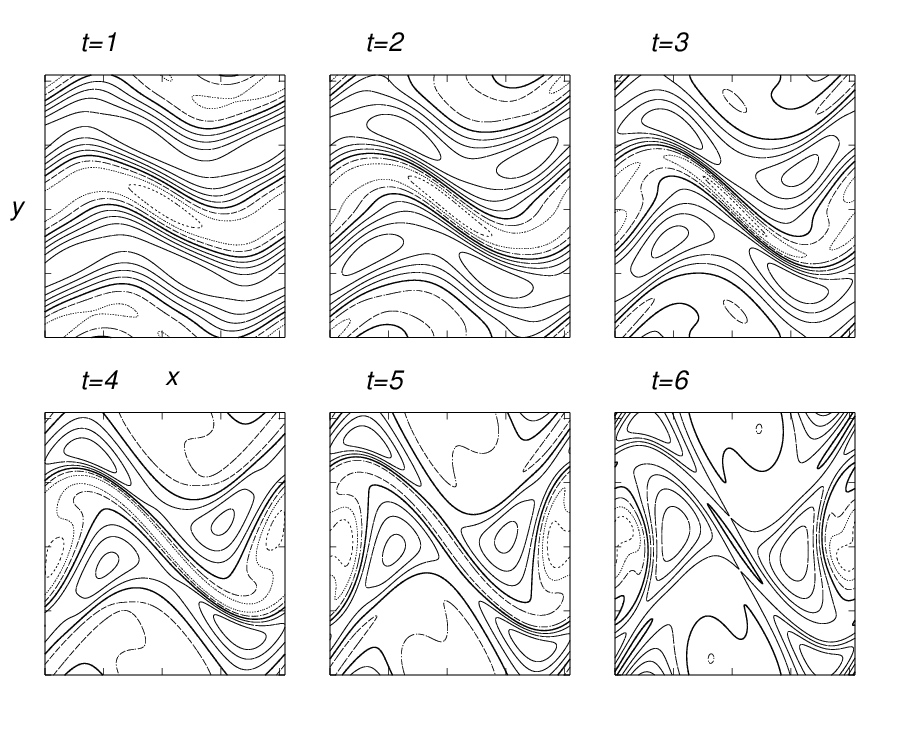}
  \caption{Contours of vorticity difference
    $\delta\omega=\left.\omega\right|_{\alpha=\frac{\pi}{32}} - \omega_{\rm NS}$ 
    near the Navier-Stokes limit, plotted as in Fig.\ref{B2N}.}
  \label{omega_diff_NS}
\end{minipage}
\end{figure}

In Fig.\ref{N2B} we show evolution of vorticity contours  of principal variation
$\left.\Omega\right|_{\alpha=0}$ near the Navier-Stokes limit, Steep gradient of
vorticity is formed near the straight line $y=-x$ at $t \lesssim 4$.
In Fig.\ref{omega_diff_NS} we show  contours of vorticity difference
$\delta\omega=\left.\omega\right|_{\alpha=\frac{\pi}{32}} - \omega_{\rm NS},$
which look similar to those in Fig.\ref{N2B} for $t \lesssim 4$.
In view of the relation $\delta\omega=\alpha\Omega$ with $\alpha=\frac{\pi}{32}$,
this again shows that the principal variation captures the deviation of vorticity
due to a small change in $\alpha$ for a while.
However, for  $t > 5$ the patterns start to differ, namely,
the principal variation starts to fail capturing $\delta \omega$ for a small, but finite $\alpha$.

\subsection{Self-similar solutions}
We make a brief remark on the self-similar solutions to (\ref{vorticity.gen}).
When we introduce dynamic scaling transformations
$$\omega(\bm{x},t)=\frac{1}{2at}\Omega(\bm{\xi},\tau), \bm{\xi}=\frac{\bm{x}}{\sqrt{2at}},
\tau=\frac{1}{2a}\log t,$$
the generalized vorticity equations are transformed to
$$\frac{\partial \Omega}{\partial \tau}+
\sin \alpha \triangle_{\bm{\xi}}  \frac{|\bm{V}|^2}{2}+\cos\alpha(\bm{V}\cdot\nabla_{\bm{\xi}} )\Omega
=\nu \triangle_{\bm{\xi}} \Omega+a\nabla_{\bm{\xi}} \cdot (\bm{\xi}\Omega).$$
It follows that their steady solutions, if any, would satisfy
\bel{ss_eq}
\sin \alpha \triangle_{\bm{\xi}} \frac{|\bm{V}|^2}{2}+\cos\alpha(\bm{V}\cdot\nabla_{\bm{\xi}} )\Omega
=\nu \triangle_{\bm{\xi}} \Omega+a\nabla_{\bm{\xi}} \cdot (\bm{\xi}\Omega),
\ee
which determine the profile of the self-similar solutions.
It is not known how to solve (\ref{ss_eq}) for general $\alpha$, except for the following cases.
When $\alpha=0,$ we have a radially symmetric solution known as the Burgers vortex
$$\Omega(\bm{\xi})=\frac{a\Gamma}{2\pi\nu}\exp\left(-\frac{a}{2\nu}|\bm{\xi}|^2 \right),$$
where $\Gamma=\int_{\mathbb{R}^2}\omega d\bm{x}$ denotes the circulation.
The corresponding (radial) stream function is given by
$$
\Psi(\bm{\xi})=\frac{\Gamma}{4\pi}{\rm Ei}\left(-\frac{a|\bm{\xi}|^2}{2\nu} \right)+c_1\log|\bm{\xi}|+c_2,
$$
where ${\rm Ei}(x)=-\int_x^\infty e^{-s}\frac{ds}{s}\;\;(x<0)$ denotes the exponential integral.
The constants are $c_1=-\frac{\Gamma}{2\pi}$ for $|\Psi(0)| < \infty$ and $c_2$ is arbitrary.

On the other hand, when $\alpha=\pi/2,$ we have $\Omega(\bm{\xi})=\triangle_{\bm{\xi}} \Psi,$
where
$$\Psi(\bm{\xi})=\log \left\{
1-\frac{K}{2\nu}\int_0^{\xi_1}\int_0^{\xi_2}\exp\left(-\frac{a}{2\nu}(\xi^2+\eta^2) \right)
\right\},$$
for a constant $K$ \cite{OV2022}.
It should be noted that the solution is linear for $\alpha=0$ because it is radial, whereas
nonlinear for $\alpha=\pi/2$. Since the superposition principle does not work,
no general solutions are known explicitly
for $0 < \alpha < \frac{\pi}{2}.$

\section{Summary and outlook}

In this paper we proposed a system of generalized fluid dynamical equations,
interpolating between the  2D Navier-Stokes and Burgers equations
and studied the system both numerically and theoretically.

We carried out numerical experiments of the generalized system.  It is  found that
the regularity property of solutions deteriorates when we increase the  parameter　from
$\alpha=0$ to $\pi/2$ and that flows remain regular for all the cases we conducted
with viscosity. We note that the worst case $\alpha =\pi/2$ is known to be integrable via
the heat kernel.

On the theoretical side we presented a formal perturbative treatment of the system, regarding the
angle as a parameter.
In particular, we derived the variational equations from either limits of the parameter $\alpha=0$
and $\pi/2$
and numerically solved the equations of principal variation, simultaneously with the Navier-Stokes
or Burgers equations.
The results are compared well
with direct numerical simulations of the cases $\alpha \gtrsim 0,$ and $\alpha \lesssim \pi/2$
for some finite time.

For a finite, but small angle parameter, this in principle offers a possibility of approximating
solutions to the 2D Navier-Stokes equations based on those of the 2D Burgers equations. However,
at the moment
the validity regions of parameter are restricted to the close vicinity of the both extreme values.　
It may be of interest to compare with other theories which attempt to solve nonlinear PDEs
based on the heat flows, such as \cite{FS1987, FHS2018}.

Also of interest is to consider a system of generalized equations  in three dimensions.
Their introduction and numerical experiments will be reported in the future.


\appendix
\section{Inviscid Burgers equation}
It is well known that the 2D Burgers equation for $\nu=0$ conserves the total kinetic energy
$$\widetilde{E}(t)= \frac{1}{2}\int_{\mathbb{R}^2}\rho |\bm{v}|^2 d\bm{x},$$
where  $\rho$ denotes the density satisfying the  continuity equation
  $$\frac{\partial \rho}{\partial t}+\nabla\cdot(\rho\bm{v})=0.$$
On the other hand, the (squared) $L^2$-norm of velocity
$$\frac{1}{2}\int_{\mathbb{R}^2} |\bm{v}|^2d\bm{x}$$
is not generally conserved  by the same equations, as noted above near (\ref{budget}).
However, for the initial condition (\ref{IC}) we studied herein, the $L^2$-norm hardly
changes in the very early stage of evolution $t \lesssim 0.5$.

\begin{figure}[htbp]
 \begin{minipage}{.49\linewidth}
  \includegraphics[scale=.3,angle=0]{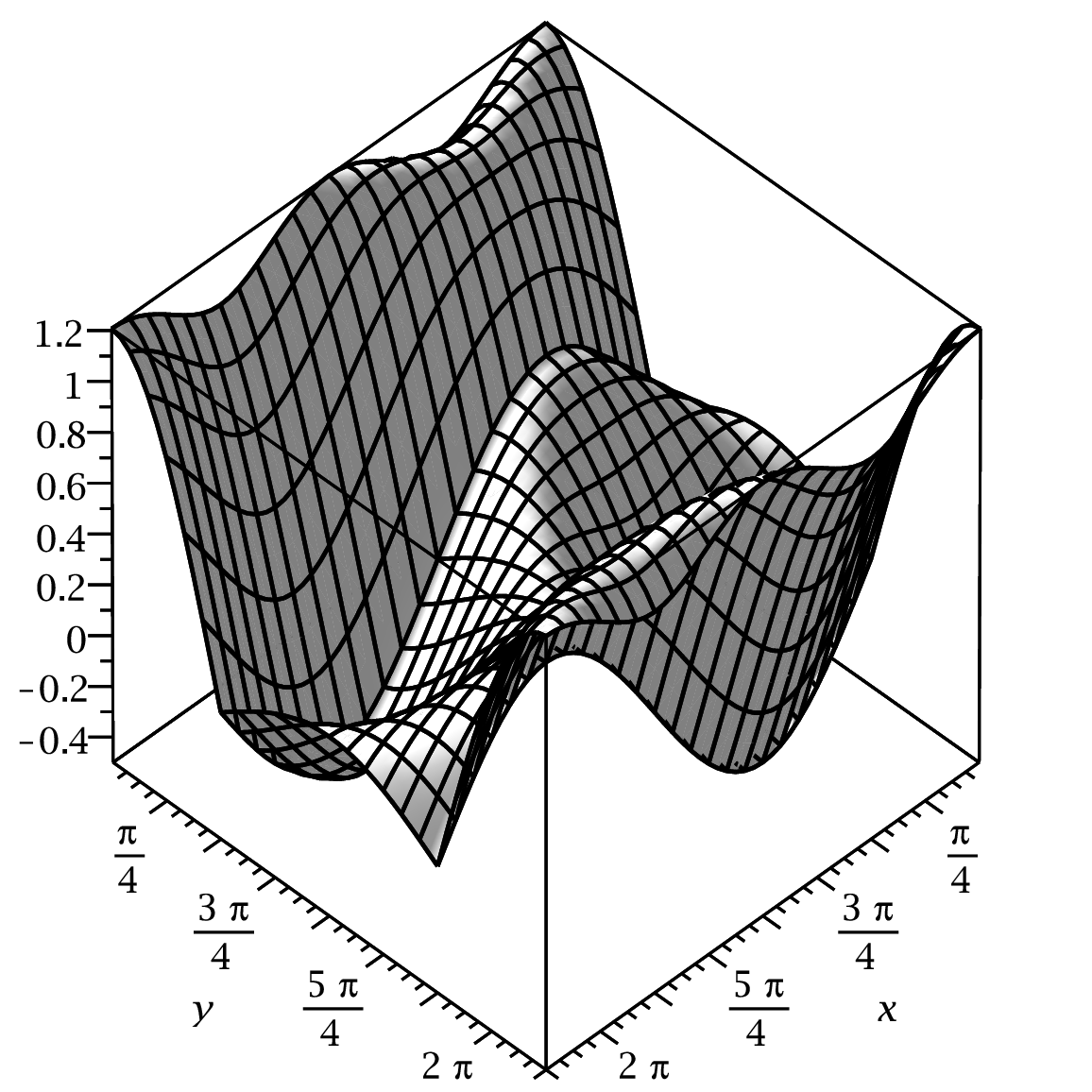}
  \caption{The eigenvalue $\lambda_+(x,y)$ of the initial velocity gradient for the Burgers
    equations.}
  \label{eigen}
\end{minipage}
\end{figure}

It is in order  to scrutinize the issue in connection with blowup.
First, it can be seen that the solution to the inviscid Burgers equations
  blows up at $t_c=2(\sqrt{2}-1)\approx 0.828$ for the initial condition in question.
  In fact, the velocity gradient tensor $W_{ij}=\partial_j u_i,\;(i,j=1,2)$ satisfies
$$
\frac{D\bm{W}}{Dt}=-\bm{W}\bm{W},
$$
where $D/Dt=\partial_t + \bm{u}\cdot\nabla$ denotes the Lagrangian derivative.
Its solution  in the Lagrangian representation is given by
$$
\bm{W}(t)=\bm{W}(0)(\bm{I}+t\bm{W}(0))^{-1},
$$
where $\bm{I}$  denotes the identity matrix.
Hence the incipient singularity takes place when the condition
$${\rm det} (\bm{I}+t\bm{W}(0))=0$$
is met for the first time at some fluid element. Clearly, the problem is reduced to that of eigenvalues
$\lambda$ of $\bm{W}(0)$.
Solving $${\rm det} (\lambda\bm{I}-\bm{W}(0))=0,$$
after some algebra we find the two eigenvalues
$$ \lambda_\pm(x,y) =\frac{\sin x \sin y+\cos y}{2}\pm \frac{|\cos y|}{2}\sqrt{1+\cos^2 x}.$$
In Fig.\ref{eigen} we plot $\lambda_+(x,y)$, which shows that the maximum value of
$|\lambda_+(x,y)|$ is given by $\lambda_+(0,0)=\frac{1+\sqrt{2}}{2}\approx 1.2.$
(The other branch $|\lambda_-(x,y)|$ gives rise to the same maximum value, hence figure omitted.)
Thus the time of breakdown is given by $t_c=\frac{2}{1+\sqrt{2}}\approx 0.83.$

\begin{figure}[htbp]
\begin{minipage}{.49\linewidth}
  \includegraphics[scale=.5,angle=0]{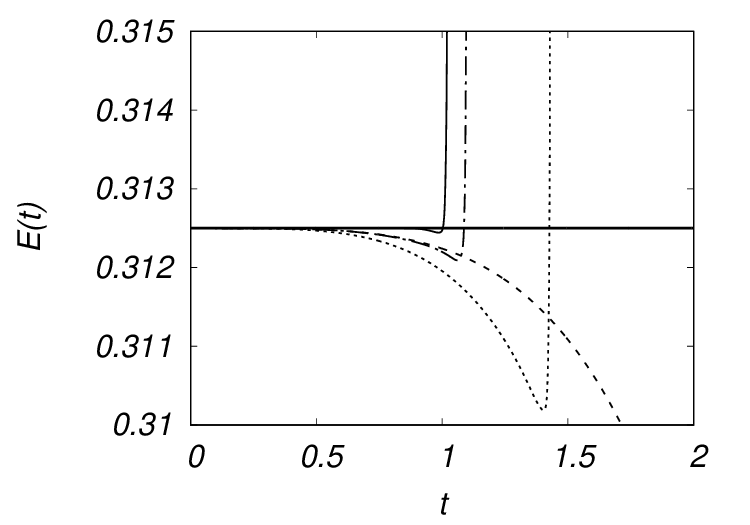}
  \caption{Time evolution of energy $E(t)$ in the inviscid case,
    for $\alpha=0$ (thick solid), $\pi/8$ (dashed), $\pi/4$ (dotted),
    $3\pi/8$ (dash-dotted) and $\pi/2$ (solid).}
    \label{ene_inv}
\end{minipage}
\begin{minipage}{.49\linewidth}
\includegraphics[scale=.5,angle=0]{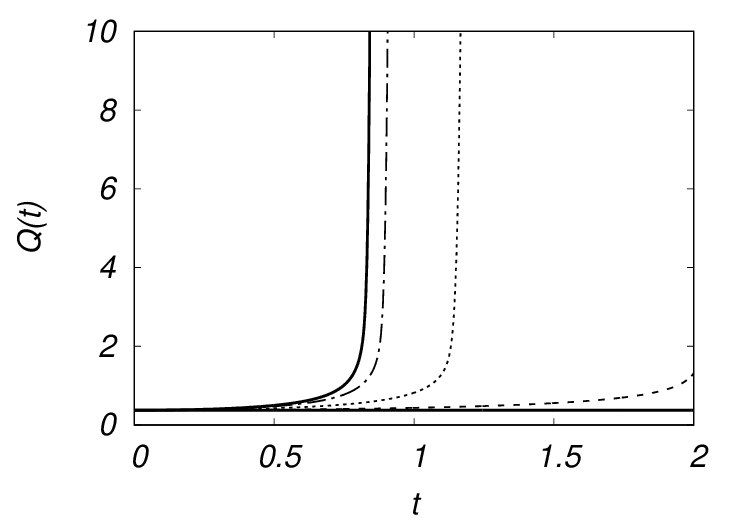}
  \caption{Time evolution of enstrophy  in the inviscid case,
    $Q(t)$ for  $\alpha=0$ (thick solid), $\pi/8$ (dashed), $\pi/4$ (dotted),
     $3\pi/8$ (dash-dotted) and $\pi/2$ (solid).} 
\label{ens_inv}
\end{minipage}
\end{figure}

In Fig.\ref{ene_inv} we show the time evolution of energy for several values of $\alpha$,
which shows breakdown as measured in this norm.
Note that the ordinate has been magnified to emphasize the virtually constant nature
of the norm for some time and the horizontal line corresponds to the 2D Euler equations.

In Fig.\ref{ens_inv} we show time evolution of enstrophy  for several values of $\alpha$.
Except for the horizontal line for the 2D Euler equations ($\alpha=0$),
all the cases are divergent at some time.
Comparing with  Fig.\ref{ene_inv}, we observe that variation of
the energy norms become noticeable when the time of (numerical) breakdown is approached.

{\bf acknowledgments}
  This work was supported by the Research Institute for Mathematical
  Sciences, an International Joint Usage/Research Center located in Kyoto
University. This work was also supported by JSPS KAKENHI Grant Number JP22K03434.


\end{document}